\documentclass[prl,aps,showpacs,twocolumn,floatfix]{revtex4}
\usepackage{epsfig} \usepackage{graphics} \usepackage{bm}
\usepackage{amssymb}
\begin{document}

\preprint{Lebed-Sepper-PRL}

\title{Non-analytical Angular Dependence of the Upper Critical Magnetic 
Field in a Quasi-One-Dimensional Superconductor}

\author{A.G. Lebed$^*$}
\author{O. Sepper}

\affiliation{Department of Physics, University of Arizona, 1118 E.
4-th Street, Tucson, AZ 85721, USA}

\begin{abstract}
We have derived the so-called gap equation, which determines the upper
critical magnetic field, perpendicular to conducting chains of a 
quasi-one-dimensional superconductor.
By analyzing this equation at low temperatures, we have found that the
calculated angular dependence of the upper critical magnetic field is 
qualitatively different than that in the so-called effective mass model.
In particular, our theory predicts a non-analytical angular dependence of
the upper critical magnetic field, $H_{c2}(0) - H_{c2}(\alpha) \sim \alpha^{3/2}$,
when magnetic field is close to some special crystallographic axis and makes
an angle $\alpha$ with it.
We discuss possible experiments on the superconductor (DMET)$_2$I$_3$
to discover this non-analytical dependence.

\end{abstract}


\maketitle


Upper critical magnetic field, which corresponds to destruction of 
superconductivity in type II superconductors, is known to be one of the most
fundamental properties of a superconducting state.
The first calculations of the upper critical magnetic field were done in
the framework of the phenomenological Ginzburg-Landau (GL) theory 
(see, for example, [1,2]) before the creation of the Bardeen-Cooper-Schrieffer 
(BCS) microscopic theory of 
superconductivity.
Later, it was shown [3] that the GL theory is a limiting case of the 
BCS theory at $T_c-T \ll T_c$ and the upper critical magnetic fields were 
calculated [4] at $T_c-T \ll T_c$ for superconductors with anisotropic
electron spectra, where $T_c$ is superconducting transition temperature
in the absence of a magnetic field.
Using the microscopic Gor'kov equations, the upper critical field was calculated
for a 3D isotropic superconductor at zero temperature [5] and at arbitrary
temperatures [6].
 As for superconductors with anisotropic electron spectra, the common belief 
 is that we can apply the results [4], obtained at $T_c - T \ll T_c$, at any 
 temperature, including $T \ll T_c$. The results [4] are usually called the effective
 mass (EM) model.
 
 The main goal of our Letter is to show that the shape and topology of the Fermi
 surface (FS) play a crucial role in determination of angular dependence of
 the upper critical magnetic field at low temperatures.
 We consider a quasi-one-dimensional (Q1D) superconductor, which is
 characterized by two open slightly corrugated sheets of 
 the FS. 
 By using the Gor'kov equations [3], we derive the so-called gap equation,
 determining the upper critical magnetic field, perpendicular to conducting 
 chains in a Q1D 
 superconductor.
 As a result, we obtain a rather complicated integral equation, which we 
 numerically solve at $T \ll T_c$.
 Our numerical analysis of this integral equation shows that the EM model 
 cannot be applied to Q1D case at $T \ll T_c$ even at 
 qualitative level.
 Our main finding is that we predict non-analytical angular dependence of 
 the upper critical magnetic field, $H_{c2}(0) - H_{c2}(\alpha) \sim
 \alpha^{3/2}$, in the case, where magnetic field is close to some special 
 crystallographic axis and makes an angle $\alpha$ 
 with it.
 This fact is in a sharp disagreement with a common belief, based on the results
 of the EM model, that $H_{c2}(0)-H_{c2}(\alpha)$ has to be proportional to 
 $\alpha^2$.
 Our second finding is that superconducting nuclei (i.e., solutions of the
 gap integral equation) are not of an exponential 
 shape. 
 We show that they decay very slowly and change their signs with 
 distance.
 It is important that the above described phenomena are novel and due
 to quasi-classical effects of an electron motion in a magnetic field along
 open sheets of the Q1D FS in a single Brillouin zone.
 They are different from quantum effects of an electron motion in the
 extended Brillouin zone, considered in Refs.[7,8]. 
 Moreover, for discovery of non-analytical angular dependence, we need
 different experimental conditions than for investigation of the so-called
 Reentrant superconductivity [7-11].
 We propose to investigate effects, suggested in the Letter, in the Q1D 
 superconductor (DMET)$_2$I$_3$, where the upper critical magnetic
 fields have been recently measured along all three principal 
 directions [12].
 It has also been pointed out [12] that superconductivity in the above 
 mentioned compound is very far from the Reentrant superconducting
 regime [7], in contrast to superconductivity in (TMTSF)$_2$X
 materials [7-11].
 
 Let us consider a superconductor with the following Q1D electron spectrum,
\begin{equation}
\delta \epsilon^{\pm}({\bf p})= \pm v_F (p_x \mp p_F) - 2t_y \cos(p_y a_y)
- 2t_z \cos(p_z a_z) ,
\end{equation}
in a magnetic field,
\begin{equation}
{\bf H} = (0,H \cos \alpha, H \sin \alpha), \ \ {\bf A} = (0,Hx \sin \alpha, -Hx \cos \alpha),
\end{equation}
perpendicular to its conducting chains.
[Here, +(-) stands for right (left) sheet of the Q1D FS (1), $t_y \gg t_z$ 
are electron hoping integrals along ${\bf a}_y$ and ${\bf a}_z$ crystallographic 
axes; $v_F$ and $p_F$ are the Fermi velocity and Fermi momentum, respectively;
$\hbar \equiv 1$.]

To determine electron wave functions in the mixed representation, $\Psi_{\epsilon}^{\pm} (x,p_y,p_z)$, where
\begin{equation}
\Psi_{\epsilon}^{\pm}(x,y,z) = \exp[i p_F x] \exp(ip_y y) \exp(ip_z z)
\Psi_{\epsilon}^{\pm} (x,p_y,p_z),
\end{equation}
we use the so-called Peierls substitution method,
$p_x \mp p_F \rightarrow -i d/dx$, $p_y \rightarrow p_y -eA_y/c$,
$p_z \rightarrow p_z -eA_z/c$.
As a result, we obtain the following electron Hamiltonian in the presence of
a magnetic field:
\begin{eqnarray}
\hat H = \mp i v_F \frac{d}{dx} - &&2t_y \cos \biggl(p_y a_y - \frac{\omega_y}{v_F}x \biggl)
\nonumber\\
- &&2t_z \cos \biggl(p_z a_z + \frac{\omega_z}{v_F}x \biggl) ,
\end{eqnarray}
where $\omega_y = ev_FH a_y \sin \alpha /c$ and 
$\omega_z = ev_FH a_z \cos \alpha /c$.

In this Letter, we ignore quantum effects of an electron motion in a magnetic 
field in the extended Brillouin zone [7-11] and use the so-called eikonal 
approximation [3].
Note that we consider the case of small angles, $\alpha \ll 1$, where $\omega_z
\gg \omega_y$, which is important for non-analytical dependence of the upper
critical field. 
As shown in Ref.[7], the quantum effects are small only at high enough 
temperature, where
\begin{equation}
T \gg T^* \simeq \frac{\omega_z}{8 \pi^2} \ln(4t_z/\omega_z)
\end{equation}
(see Eq.(6) of Ref.[7]).
Under condition (5), we can linearize the Hamiltonian (4) with respect to a magnetic
field,
\begin{eqnarray}
\hat H = \mp i v_F \frac{d}{dx} - &&2t_y \cos (p_y a_y) - 2 t_y \frac{\omega_y x}{v_F}
\sin (p_y a_y)
\nonumber\\
- &&2t_z \cos (p_z a_z) + 2 t_z \frac{\omega_z x}{v_F}
\sin (p_z a_z) .
\end{eqnarray}
It is important that the corresponding Schrodinger equation for wave functions 
in the mixed representation,
\begin{equation}
\hat H \Psi_{\epsilon} (x, p_y, p_z) = \delta \epsilon \Psi_{\epsilon} (x, p_y, p_z) ,
\end{equation}
can be exactly solved:
\begin{eqnarray}
\Psi_{\epsilon} (x, p_y, p_z) = \exp(\pm i \delta \epsilon x / v_F) 
&&\exp [ \pm i \phi_y(p_y,x)] 
\nonumber\\
&&\exp [ \pm i \phi_z(p_z,x)],
\end{eqnarray} 
where
\begin{eqnarray}
&&\phi_y(p_y,x) = \frac{2t_y}{v_F} \cos(p_y a_y) x + \frac{t_y \omega_y}{v^2_F} \sin(p_ya_y) x^2 ,
\nonumber\\
&&\phi_z(p_z,x) = \frac{2t_z}{v_F} \cos(p_z a_z) x - \frac{t_z \omega_z}{v^2_F} 
\sin(p_z a_z) x^2 .
\end{eqnarray} 

Since the electron spectrum and wave functions are known, the corresponding
finite temperatures Green functions can be determined by means of the standard 
procedure [13]:
\begin{eqnarray}
&&G^{\pm}_{i \omega_n} ({\bf r},{\bf r_1}) = \frac{-i \ sgn(\omega_n)}{v_F}
\sum_{p_y,p_z} \exp[\pm i p_F (x-x_1)] 
\nonumber\\
&&\times \exp[i p_y(y-y_1)] \exp[i p_z(z-z_1)] \exp \biggl[ \frac{\mp \omega_n (x-x_1)}{v_F}
\biggl]
\nonumber\\
&&\times \exp [ \pm i 2 t_y  \cos(p_y a_y)(x-x_1) / v_F] 
\nonumber\\
&&\times \exp [ \pm i 2t_z \cos(p_z a_z)(x-x_1) / v_F]
\nonumber\\
&&\times \exp[ \pm i t_y \omega_y \sin(p_y a_y) (x^2-x^2_1)/v^2_F]
\nonumber\\
&&\times \exp[ \mp i t_z \omega_z \sin(p_z a_z) (x^2-x^2_1)/v^2_F] .
\end{eqnarray}

The so-called gap equation, determining superconducting transition temperature 
in the presence of the magnetic field (2), can be derived by using the Gor'kov
equations for non-uniform superconductivity [14].
As a result, we obtain:
\begin{eqnarray}
&&\Delta(x) = \frac{g}{2} \int_{|x-x_1| > d} \frac{2 \pi T d x_1}{v_F \sinh (2 \pi T |x-x_1|/v_F)}
\nonumber\\
&&J_0 \biggl[ \frac{2 t_y \omega_y}{v^2_F} (x^2-x^2_1) \biggl]
J_0 \biggl[ \frac{2 t_z \omega_z}{v^2_F} (x^2-x^2_1) \biggl] \Delta(x_1) ,
\end{eqnarray} 
where $g$ is an effective electron coupling constant, $d$ is a cutoff 
distance [15].
Here, we rewrite Eq.(11) in more convenient way:
\begin{eqnarray}
&&\Delta(x) = \frac{g}{2} \int_{|z| > d} \frac{2 \pi T d z}{v_F \sinh (2 \pi T |z|/v_F)}
\nonumber\\
&&J_0 \biggl[ \frac{2 t_y \omega_y}{v^2_F} z (z+2x) \biggl]
J_0 \biggl[ \frac{2 t_z \omega_z}{v^2_F} z (z+2x) \biggl] \Delta(x+z) .
\end{eqnarray} 
[Note that the Pauli paramagnetic spin-splitting effects are ignored
in all equations above, which means that the upper critical magnetic field 
is supposed to be much smaller than the so-called Clogston-Chandrasekhar paramagnetic limit [17].
Such situation, for example, is experimentally realized in the Q1D superconductor 
(DMET)$_2$I$_3$ [12].]

Let us determine the GL slope of the upper critical magnetic field in the 
vicinity of superconducting transition temperature.
To achieve this goal, we need to take into account that in the GL region, 
$(T_c-T) \ll T_c$,
$v_F /2 \pi T_c \ll v_F / \sqrt{t_y \omega_y}, v_F / \sqrt{t_z \omega_z}$.
In this case, we can expand the integral equation (12) in terms of small
parameter, $|z| \sim v_F / 2 \pi T_c$.
As a result of such expansion procedure, we obtain the following differential
equation:
\begin{eqnarray}
&&\biggl[ -\frac{d^2 \Delta(x)}{dx^2} + x^2 \frac{8 (t^2_y \omega^2_y +t^2_z \omega^2_z)}{v^4_F}  \Delta(x)   \biggl]      \int^{\infty}_0 
\frac{\pi  T_c z^2 dz}{v_F \sinh (2 \pi T_c  / v_F)}
\nonumber\\
&&+\biggl[ \frac{1}{g} -  \int^{\infty}_d  \frac{2 \pi T dz}{v_F \sinh ( 2 \pi T z / v_F)} \biggl]\Delta (x) = 0.
\end{eqnarray}
If we take into account that
\begin{equation}
 \frac{1}{g} =  \int^{\infty}_d  \frac{2 \pi T_c dz}{v_F \sinh ( 2 \pi T_c z / v_F)}  ,
\end{equation}
then we can rewrite Eq.(13) in the following way:
\begin{eqnarray}
- \xi^2_x \frac{d^2 \Delta(x)}{dx^2} + &&\biggl( \frac{2 \pi H}{\phi_0} \biggl)^2 (\xi_y^2 \sin^2 \alpha + \xi_z^2 \cos^2 \alpha) \ x^2 
\Delta(x) 
\nonumber\\
&&- \tau \Delta(x) = 0 ,
\end{eqnarray}
where
\begin{eqnarray}
\xi^2_x = \frac{7 \zeta(3) v^2_F}{16 (\pi T_c)^2}  , \ \xi^2_y = \frac{7 \zeta(3) t^2_y a^2_y}{8 (\pi T_c)^2} , &&\xi^2_z = \frac{7 \zeta(3) t^2_z a^2_z}{8 (\pi T_c)^2} ,
\nonumber\\
 &&\tau = \frac{T_c-T}{T_c}, 
\end{eqnarray}
[Here $\phi_0 = \pi \hbar c /e$ is the flux quantum, $\xi_x$, $\xi_y$, and $\xi_z$ are 
the coherence lengths along ${\bf a}_x$, ${\bf a}_y$, and ${\bf a}_z$ axes,
correspondingly.]
Note that above we use the following relationship:
\begin{equation}
\int^{\infty}_0 \frac{z^2 dz}{\sinh(z)} = \frac{7 \zeta(3)}{2},
\end{equation}
where $\zeta(n)$ is the Reimann zeta function [18].
To find the GL slope of the upper critical magnetic field, perpendicular the
conducting chains, we need to determine the lowest energy level of the 
Schrodinger-like GL equation (15).
As a result, we obtain
\begin{equation}
\frac{1}{H^2_{c2} (\alpha)} = \frac{\sin^2 \alpha}{H^2_{c2}(\pi/2)} + \frac{\cos^2 \alpha}{H^2_{c2}(0)} ,
\end{equation} 
where
\begin{eqnarray}
&&H_{c2}(0,T) = \frac{\phi_0}{2 \pi \xi_x \xi_z} \biggl( \frac{T_c-T}{T_c} \biggl) =
\frac{4 \sqrt{2} \pi^2 c T^2_c}{7 \zeta(3) e v_F t_z a_z} \biggl( \frac{T_c-T}{T_c}
\biggl) ,
\nonumber\\
&&H_{c2} \biggl( \frac{\pi}{2} , T \biggl) = \frac{\phi_0}{2 \pi \xi_x \xi_y} \biggl( \frac{T_c-T}{T_c} \biggl) =
\frac{4 \sqrt{2} \pi^2 c  T^2_c}{7 \zeta(3) e v_F t_y a_y} \biggl( \frac{T_c-T}{T_c}
\biggl).
\end{eqnarray}
[Note that Eq.(18) is usually called EM model and applied to fit the 
experimental upper critical magnetic fields at any temperature, including 
$T \ll T_c$. 
On the other hand, we pay attention that Eqs.(13)-(19) are derived under the
GL condition $(T-T_c) \ll T_c$, which is equivalent to the following two conditions:
$T^2_c \gg t_z \omega_z$ and $T^2_c \gg t_y \omega_y$. 
It is important that the latter inequalities can be rewritten as: $H \ll H_{c2}(0, T=0)$ 
and $H \sin \alpha \ll H_{c2}(\pi/2, T=0)$.]

Below, we consider the gap equation (12) at low temperature, $T_c \gg T \gg T^*$, 
where we can formally employ $T=0$ in Eq.(12):
\begin{eqnarray}
&&\Delta(x) = \frac{g}{2} \int^{\infty}_{\sqrt{2 \omega_0 t_z} d /v_F} \frac{d z}{z} \biggl\{
J_0[\beta \sin (\alpha) z (2x+z)]  
\nonumber\\
&&\times J_0[ \cos (\alpha) z (2x+z)] 
 \Delta (x+z)
+ J_0[\beta \sin (\alpha) z (2x-z)] 
\nonumber\\ 
&&\times J_0[ \cos (\alpha) z (2x-z)] \Delta (x-z) \biggl\} ,
\end{eqnarray} 
where $\beta = t_y a_y / t_z a_z$, 
$\omega_0 = ev_FHa_z/c$.
It is important that the effective electron coupling constant, $g$, and cutoff 
distance, $d$, can be eliminated from Eq.(20) by using the following 
relationship:
\begin{equation}
 \frac{1}{g} =  \int^{\infty}_{2 \pi T_c d/v_F}  \frac{dz}{v_F \sinh (z)} ,
\end{equation}
which is a result of Eq.(14).

\begin{figure}[t]
\begin{center}
\epsfig{file=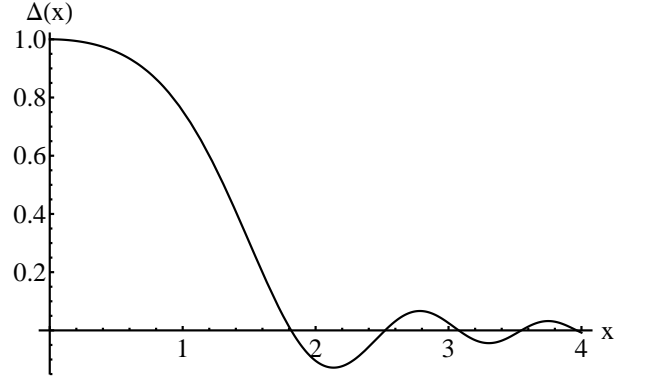,width=8cm} \caption{\label{fig1}
A typical solution of Eqs.(20),(21) is shown for $\alpha=0$. 
Note that it is an oscillatory slow decaying function of coordinate $x$,
in contrast to exponential solution of Eq.(15) in the EM model [4].}
\end{center}
\end{figure}

\begin{figure}[t]
\begin{center}
\epsfig{file=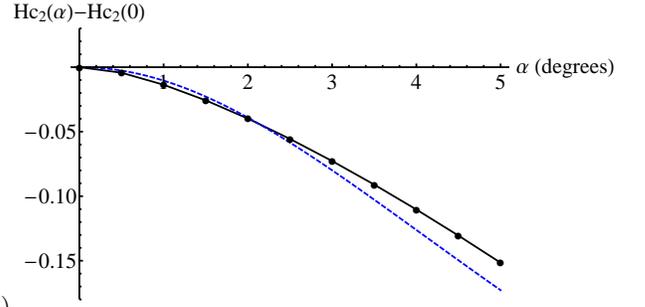,width=8cm} \caption{\label{fig2}
Angular dependence of the upper critical magnetic field at $T=0$,
$H_{c2}(\alpha)-H_{c2}(0)$, calculated from Eqs.(20),(21),
is shown by dots, which are well fitted by function $-\alpha^{3/2}$
(solid line).
The EM model result (18), where $H_{c2}(\alpha)-H_{c2}(0) \sim - \alpha^2$, 
is shown by dashed line for a comparison.}
\end{center}
\end{figure}

\begin{figure}[t]
\begin{center}
\epsfig{file=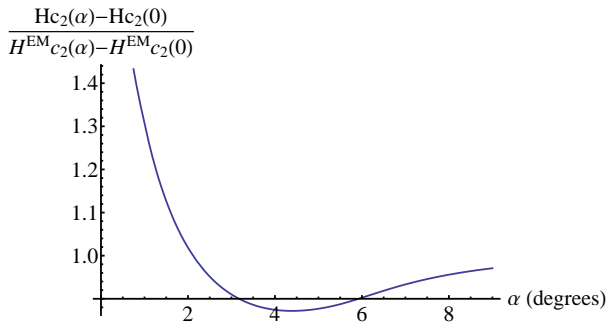,width=8cm} \caption{\label{fig3}
Calculated by means of Eqs.(20),(21) and normalized angular dependence 
of the upper critical magnetic field (see the main text).}
\end{center}
\end{figure}

Note that experimental value of the parameter $\beta$ in (DMET)$_2$I$_3$
superconductor is estimated as 
$\beta \simeq 10$ [12].
Below, we analyze Eqs.(20),(21) numerically by solving the gap integral equation 
(20) under the condition (21) for $\beta =10$.
Let us first consider the case $\alpha=0$, where magnetic field is applied
along ${\bf a}_y$ axis.
A typical solution of Eq.(20), which is called superconducting nucleus, in this case 
is shown in Fig.1.
As seen from Fig.1, in our case superconducting nucleus changes its sign
and slowly decays in space, in contrast to the results of the EM 
model [4].
Note that at $\alpha \neq 0$ solutions of Eqs.(20),(21) become more complicated,
but they retain the above mentioned unusual 
properties.
In Fig.2, we show the calculated angular dependence $H_{c2}(\alpha) - H_{c2}(0)$
and its fit by function $- \alpha^{3/2}$.
Note that the agreement between the calculated angular dependence and function
$- \alpha^{3/2}$ is very good.
For comparison, we also show dependence (18), expected in the EM model,
where $H_{c2}(\alpha) - H_{c2}(0) \sim - \alpha^2$ [19].

\begin{figure}[t]
\begin{center}
\epsfig{file=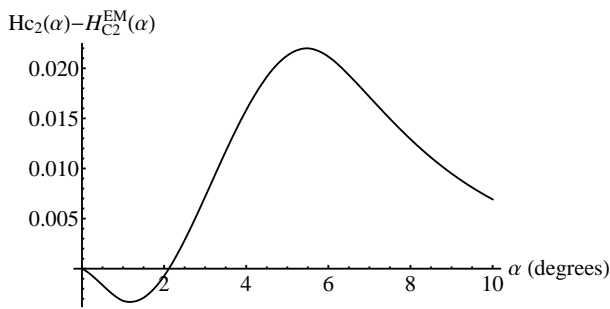,width=8cm} \caption{\label{fig4}
Calculated difference of angular dependence of the upper critical 
magnetic field, given by Eqs.(20),(21), and that, given by the EM model (18), $H_{c2}(\alpha)-H^{EM}_{c2}(\alpha)$.
}
\end{center}
\end{figure}

In Fig.3, we plot the calculated angular dependence of the upper critical
magnetic field, normalized on the corresponding result (18) of the EM model,
$[H_{c2}(\alpha) - H_{c2}(0)] / [H^{EM}_{c2}(\alpha)-H^{EM}_{c2}(0)]$.
As it follows from Fig.3, the maximum deviations from the EM model occur
at low angles and in the vicinity of some angle 
$\alpha \simeq 5^o$.
At low angles, the calculated in the Letter upper critical magnetic field exhibits
different angular dependence than that in the EM model (18),
as discussed above.
To clarify nature of the minimum in Fig.3 at $\alpha \simeq 5^o$, we 
plot the difference, $H_{c2}(\alpha) - H^{EM}_{c2}(\alpha)$,
in Fig.4.
As seen from Fig.4, the maximum difference corresponds to $\alpha \simeq 5.6^o$ -
angle, which we relate to the following theoretical value:
\begin{equation}
\alpha^* = \arctan(1/\beta) = \arctan(1/10) \simeq 5.71^o .
\end{equation}
Note that, under condition (22), both Bessel functions in Eq.(20) have
the same arguments and some kind of resonance appears.
We suggest to measure experimentally the position of the peak in the
angular dependence $H_{c2}(\alpha) - H^{EM}_{c2}(\alpha)$ to carefully determine 
the ratio $\beta = t_y a_y / t_z a_z$ from Eq. (22).

To summarize, we have shown that the EM model [4] is not adequate to describe
the upper critical magnetic field in superconductors with anisotropic electron
spectra at low temperatures.  
For the case of a Q1D superconductor, we have found non-analytical angular
dependence of the upper critical magnetic field, $H_{c2}(\alpha)-H_{c2}(0) \sim - \alpha^{3/2}$, where a magnetic field is perpendicular to conducting axis, ${\bf a}_x$,
and makes angle $\alpha$ with axis ${\bf a}_y$.
In addition, some angular resonance is predicted for "magic" direction of a magnetic
field (22).
We suggest to test the above mentioned predictions of the Letter on the Q1D
superconductor (DMET)$_2$I$_3$, where the upper critical magnetic fields along the
main crystallographic axes have been recently measured [12].
In our opinion, unconventional shapes of superconducting nuclei as well as the non-analytical angular behavior of the upper critical field, found in the Letter, may reflect 
the existence of unusual vortex lattice in Q1D superconductors.
Therefore, we also suggest experimental studies of the vortex lattice at magnetic
fields, corresponding to small values of angle $\alpha$ in Eq.(2).

Let us prove that the (DMET)$_2$I$_3$ superconductor satisfies the condition of
a validity of our theory,
\begin{equation}
T_c \gg T \gg T^* \ ,
\end{equation}
at experimentally used lowest temperature, $T \simeq  0.05 \ K$, where $T^*$ is 
given by Eq.(5) and $T_c = 0.5 \ K$ [12].
If we take from Ref.[12] the typical experimental values, $H^y_{c2} = 0.2 \ T$, 
$v_F \simeq 0.4 \times 10^7 cm/s$, $a_z = 15.8 A$, $t_z \simeq 1 \ K$, we obtain 
$T^* \simeq 0.006 \ K$.
Therefore, we conclude that the suggested in the Letter theory is applicable to
the superconductor (DMET)$_2$I$_3$ at the lowest experimental temperature
[12].
Note that, for neglecting the quantum corrections [7,8] and, thus, the Reentrant Superconductivity effects  [7-11], it is also important that $4 t_z / \omega_z \simeq 
27 \gg 1$, as has been already mentioned in Ref.[12].

We point out that in a geometry, considered in the Letter, experiments were performed in Ref.[20] in the superconductor (TMTSF)$_2$ClO$_4$ in low magnetic fields, $H \ll H_{c2}$, to demonstrate another phenomenon - the so-called 
lock-in effect. 
To avoid lock-in effect [20], the experiments, suggested by us, have to be performed at magnetic fields, which satisfy the condition $H^z \simeq H^y_{c2} \sin \alpha \gg H^z_{c1}$ [20].
Although $H^z_{c1}$ is not known in the superconductor (DMET)$_2$I$_3$, it is 
clear that in this typical type-II superconductor $H^z_{c1} \ll H^z_{c2} = 0.02 \ T$,
which shows that the above mentioned condition is presumably satisfied.

We are thankful to N.N. Bagmet for useful discussions. This work was
supported by the NSF under Grant No DMR-0705986 and Grant No
DMR-1104512.

$^*$Also at the Landau Institute for Theoretical Physics,
2 Kosygina Street, Moscow 117334, Russia.


\begin{references}
\bibitem{Lebed-6}
Note that the integral equation (11) is neither the Ginzburg-Landau
equation nor the Lawrence-Doniach one.
In this content, we stress that the condition (5) (which is necessary for the
eikonal approximation) does not necessarily lead to the Ginzburg-Landau or 
Lawrence-Doniach equations, in contrast
to the statement of Ref.[9].
Indeed, for the Lawrence-Doniach equation one needs that 
$\xi_z \ll a_z / \sqrt{2}$ [16], which is equivalent to $T_c \gg t_z$, 
which is not satisfied in our case.
For the Ginzburg-Landau equation, as we show below, one needs that
$T^2_c \gg t_z \omega_z, t_y \omega_y$ - conditions different from 
Eq.(5).
\bibitem{Lebed-7}
To the best of our knowledge, Eqs. (20),(21) cannot be analytically solved. 
Nevertheless, we are currently investigating if it is possible or not to confirm 
analytically the  angular dependence $H_{c2}(0)-H_{c2}(\alpha) \sim \alpha^{3/2}$,
numerically calculated in the Letter.
\end{references}
\end{document}